\documentclass[
preprint,
superscriptaddress,
 amsmath,amssymb,
 aps,showkeys,
]{revtex4}

\usepackage{graphicx}
\usepackage{dcolumn}
\usepackage{bm}
\usepackage{hyperref}

\usepackage{xcolor}
\newcommand{\bt}[1]{\textcolor{blue}{#1}}                           
\newcommand{\rt}[1]{\textcolor{red}{#1}}                            
\usepackage{soul}
\usepackage[normalem]{ulem}
\newcommand{\rsmath}[1]{\bgroup\markoverwith{\textcolor{red}{\rule[0.5ex]{2pt}{0.4pt}}}\ULon {\textcolor{red}{#1}}}                                           

\usepackage{eso-pic}
\AddToShipoutPictureBG*{%
  \AtPageUpperLeft{%
    \hspace{\paperwidth}%
    \raisebox{-\baselineskip}{%
      \makebox[0pt][r]{\bt{Blue texts are the deletions}, and \rt{red texts are the additions} compared with the previous manuscript.}
}}}%

\begin{document}

\preprint{APS/123-QED}

\title{Random organization and non-equilibrium hyperuniform fluids on a sphere}

\author{Yusheng Lei}
\affiliation{School of Chemistry, Chemical Engineering and Biotechnology, Nanyang Technological University, 62 Nanyang Drive, 637459, Singapore}%
\author{Ning Zheng}
\email{ningzheng@bit.edu.cn}
\affiliation{School of Physics, Beijing Institute of Technology, Beijing 100081, China}%
\author{Ran Ni}
\email{r.ni@ntu.edu.sg}
\affiliation{School of Chemistry, Chemical Engineering and Biotechnology, Nanyang Technological University, 62 Nanyang Drive, 637459, Singapore}%

\date{\today}

\begin{abstract}
Random organizing hyperuniform fluid induced by reciprocal activation is a non-equilibrium fluid with  vanishing density fluctuations at large length scales like crystals. Here we extend this new state of matter to a closed manifold, namely a spherical surface. We find that the random organization on a spherical surface behaves similar to that in two dimensional Euclidean space, and the absorbing transition on a sphere also belongs to the conserved directed percolation universality class. Moreover, the reciprocal activation can also induce a non-equilibrium hyperuniform fluid on a sphere. The spherical structure factor at the absorbing transition and the non-equilibrium hyperuniform fluid phases are scaled as $S(l \rightarrow 0) \sim (l/R)^{0.45}$ and $S(l \rightarrow 0) \sim l(l+1)/R^2$, respectively, which are both hyperuniform according to the definition of hyperuniformity on a sphere with $l$ the wave number and $R$ the radius of the spherical surface. 
We also consider the impact of inertia in realistic hyperuniform fluids, and it is found only adding an extra length-scale, above which hyperuniform scaling appears. Our finding suggests a new method for creating non-equilibrium hyperuniform fluids on closed manifolds to avoid boundary effects.
\end{abstract}


\maketitle


Disordered hyperuniform structures are an exotic state  of matter having vanishing long-wavelength density fluctuations with the structure factor $S(q \rightarrow 0)=0$ \cite{torquato2003local}. In past decades, hyperuniform structures have been found in various systems, including quasi-crystals \cite{ouguz2017hyperuniformity}, perturbed lattices \cite{kim2018effect}, binary hard-disks plasmas \cite{lomba2017disordered}, perfect glasses \cite{zhang2016perfect}, jammed structures \cite{donev2005unexpected, ricouvier2017optimizing}, avian photoreceptor patterns \cite{jiao2014avian}, biological tissues \cite{zheng2020hyperuniformity} as well as many non-equilibrium dynamic systems like driven emulsions \cite{weijs2015emergent}, driven granular systems \cite{castillo2019hyperuniform}, sheared colloids \cite{Mitra_2021, hexner2015hyperuniformity, tjhung2015hyperuniform, hexner2017noise, hexner2017enhanced, wang2018hyperuniformity, wilken2020hyperuniform} and chiral active particles \cite{lei2019nonequilibrium,lei2019hydrodynamics, zhang2022hyperuniform, huang2021circular}. Hyperuniform systems have profound applications in various fields such as enhancing super-selectivity in multivalent sensing systems \cite{xia2022receptor} and being used as photonic materials with non-trivial band gaps \cite{florescu2009designer,man2013isotropic} or transparent materials \cite{leseur2016high}.

By definition, the direct correlation function in disordered hyperuniform structures is long-ranged, and for self-organized hyperuniform states, the boundary effect is not avoidable in Euclidean spaces.
Recently, the concept of hyperuniformity has been also generalized to curved spaces mathematically~\cite{brauchart2019hyperuniform,meyra2019hyperuniformity,bovzivc2019spherical}. 
In a curved closed manifold, in principle, the boundary effect can be avoided, while the effect of topology of space on  self-organized hyperuniform states remains unknown. 
To this end, we investigate the random organization, which has been shown to form dynamic hyperuniform states in Euclidean spaces, on a closed manifold, i.e., a spherical surface.
We find that similar to the situation in Euclidean spaces, the absorbing transition on a spherical surface also belongs to the  conserved directed percolation (CDP) universality class. Both the critical hyperuniform state and the non-equilibrium hyperuniform fluid are found in absorbing transition point and the active state with momentum conserved activation, respectively, and the scaling of the spherical structure factor obtained in computer simulations agrees quantitatively with the theoretical prediction.

\begin{figure}[t]
    \centering
    \includegraphics[width=0.45\textwidth]{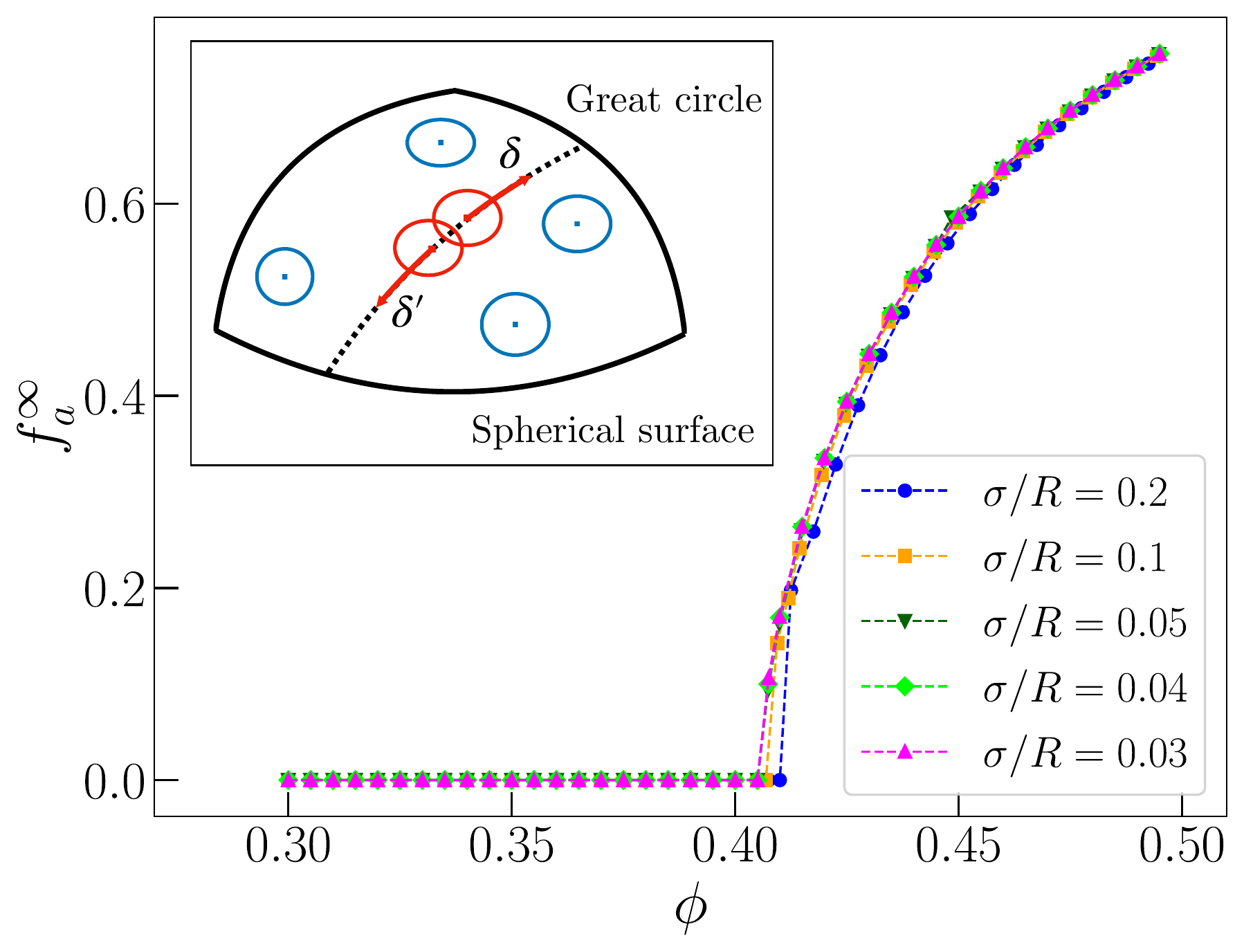}
    \caption{Steady state active particle fraction $f_a^\infty$ for various packing fraction $\phi$ at different spherical surface size $\sigma/R$, {the results are averaged from 100 configurations obtained by independent simulations from random initial configurations;} Inset: Schematic of the random organization model on a sphere. }
    \label{Fig.1}
\end{figure}
In analogy to the random organization in 2D~\cite{hexner2015hyperuniformity,tjhung2015hyperuniform,hexner2017noise}, we generalize it onto a spherical surface by performing similar local random organizing dynamics. As shown in the inset of Fig. 1, at each simulation step, particles are checked for overlap, and the overlapped particles are labelled as active particles (red circles in the figure), otherwise passive particles (blue circles in the figure). {Random displacements $\delta$ and $\delta'$ of the same magnitude but along opposite directions are assigned to a pair of overlapped particles~\cite{note1}, ensuring reciprocal particle activation.} Here the curvature effect of space can be described by the particle-sphere size ratio $\sigma/R$, where $\sigma$ and $R$ are the particle diameter and the radius of the spherical surface, respectively. The packing fraction of the system is $\phi =N\sigma^2/(16R^2)$, and the maximum displacement for active particles is fixed as $\delta_{\max} = \sigma$.

Similar to the random organization in Euclidean spaces, we choose the steady-state active particle fraction, $f_{a}^\infty$, as the order parameter of the system. As shown in Fig.~\ref{Fig.1}, we plot $f_{a}^\infty$as a function of the packing fraction, $\phi$, on the sphere surface of different radius $R$.
We can see that different from the absorbing transition in 2D, with increasing $\phi$ on a spherical surface $f_{a}^\infty$ jumps from zero to a finite number at the transition point $\phi_c$, which is due to the finite nature of the system on a spherical surface. With increasing $R$, the transition becomes smoother, and $\phi_c \rightarrow 0.405$ with $R \rightarrow \infty$.


\begin{figure*}
    \centering
    \includegraphics[width=0.95\textwidth]{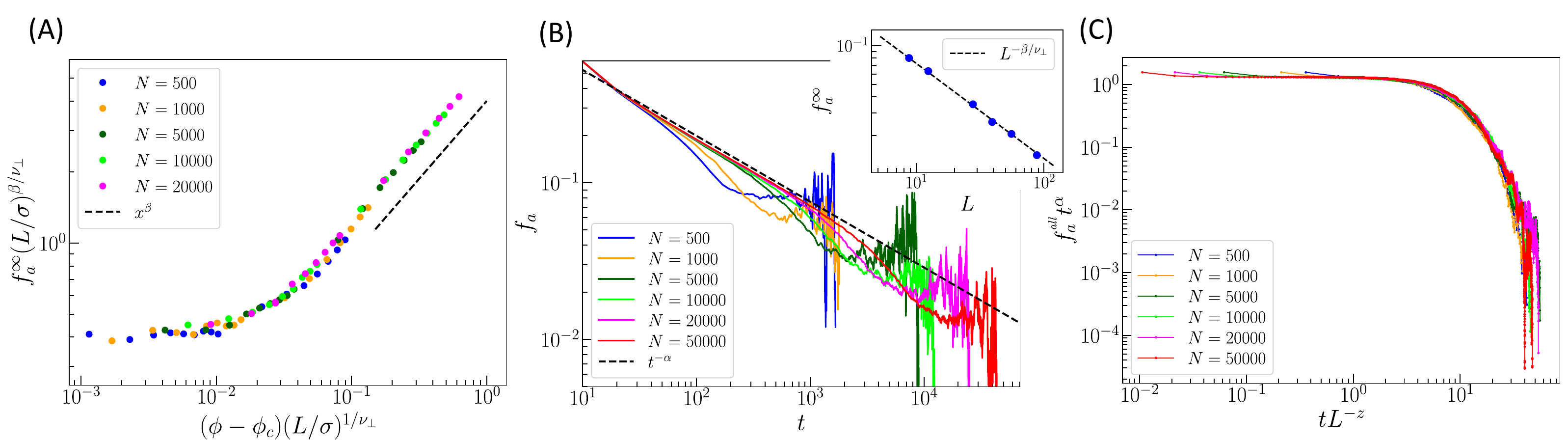}
    \caption{Finite-size scaling analysis of critical phenomena for random organization on a spherical surface. (A) Collapse of active particle fraction $f^\infty_a(\Delta \phi)$ at the critical packing fraction $\phi_c=0.405(1)$ of different system sizes $L$ through the re-scaling using Eq.~\ref{eq1} with critical exponents $\beta=0.66$, $\nu_\perp=0.89$. (B) The time evolution of the active particle fraction $f_a$ at the critical point $\phi_c$ of different system sizes, where the dash line indicates a power-law decay, $f_a \sim t^{-\alpha}$, with $\alpha=0.42$. Inset: The saturated active particle fraction at the critical point $f^\infty_a(\Delta \phi=0)$ as a function of system size $L$, where the slope of the dash line is $-\beta/\nu_\perp =-0.74$. (C) Collapse of overall active particle fraction $f_a^{all}(t)$ for systems of different  sizes at the critical point $\phi_c$ through re-scaling using Eq.~\ref{eq2} with $\alpha=0.42$, $z=1.53$. {Each data point is averaged from 1000 configurations obtained by independent experiments with random intitial configurations.}}
    \label{Fig.2}
\end{figure*}

To understand the critical behaviour of the random organization on a spherical surface, we perform a finite size scaling analysis for the absorbing transition as shown in Fig.~\ref{Fig.2}. 
For systems approaching the critical point, as shown in Fig.~\ref{Fig.2}A, the saturated active particle fraction, $f_a^\infty(\Delta\phi, L)$, satisfies the universal finite-size scaling form as described in~\cite{rossi2000universality,lei2019hydrodynamics}
\begin{equation}\label{eq1}
f^\infty_a(\Delta \phi, L) = L^{-\beta/\nu_\perp} \mathcal{G} \left( L^{1/\nu_\perp} \Delta \phi \right),
\end{equation}
where $\Delta\phi=(\phi-\phi_c)$, and $L$ is the size of the system. Here we use the radius of the spherical surface to indicate the size of the system. Therefore, $L \propto N^{1/d}$, and $d=2$ in our system. $\mathcal{G}$ is a universal scaling function with $\mathcal{G}(x)\sim x^\beta$ for large $x$, where $\beta=0.66$ and $\nu_\perp=0.89$ are the critical exponents obtained in the scaling of the spatial correlation length, $\xi_\perp \sim |\Delta\phi|^{-\nu_\perp}$, in the system.
For a finite system at the critical point $(\Delta\phi=0)$, as shown in Fig.~\ref{Fig.2}B, the evolution of the active particle fraction, $f_{a}(t)$, from random initial configurations follows a power-law decay, $f_{a}(t)\sim t^{-\alpha}$, with $\alpha=0.42$, until reaching the saturated value $f_a^\infty(\Delta\phi=0)$. The saturated active particle fraction at the critical point obeys a power-law scaling, $f_a^\infty(\Delta\phi=0)\sim L^{-\beta/\nu_\perp}$, for the system size $L$, where $\beta/\nu_\perp \approx 0.74$ can be obtained from the power-law fit in inset of Fig.~\ref{Fig.2}B.

Furthermore, as shown in Fig.~\ref{Fig.2}C, one can also consider the overall active particle fraction, $f_a^{all}$, which is averaged over both surviving and non-surviving trials. $f_a^{all}$ satisfies the finite-size scaling relationship at the critical point~\cite{rossi2000universality,lei2019hydrodynamics}
\begin{equation}\label{eq2}
f_a^{all}(t) = t^{-\alpha} \mathcal{F}\left( t L^{-z} \right),
\end{equation}
where the dynamic exponent $z=\nu_\parallel/\nu_\perp \approx 0.53$, and $\nu_\parallel$ is the scaling exponent of the temporal correlation length with $\xi_\parallel\sim|\Delta \rho|^{-\nu_\parallel}$ and $\mathcal{F}(\cdot)$ a universal scaling function. 

To understand the influence of space topology on the absorbing transition, we also perform the finite size scaling for the corresponding system in 2D, and the obtained critical exponents are compared with the random organization on a sphere in Table~\ref{tab:Table.1}. We can see that the obtained critical exponents of random organization on a sphere are very close to those in a flat 2D system, and this suggests that random organization on a sphere also belongs to the CDP universality class as in 2D~\cite{rossi2000universality,lei2019hydrodynamics,henkel2008non}. {It is known that the universality class of a phase transition mainly depends on the local rule and major conservation laws of the system. For random organization in Euclidean spaces, it was found that by only changing the local rule of introducing center-of-mass conserving noise does not change the universality class of the absorbing transition, as it does not break the particle number conservation law,~\cite{hexner2015hyperuniformity,hexner2017noise} and here our results suggest that the change of space topology may not alter the critical point universality class as well, if it does not break the major conservation law of the system.}

\begin{table}[h]
\caption{\label{tab:Table.1} Comparison of critical exponents obtained in random organization in a 2D flat plane and a spherical surface.}
\begin{ruledtabular}
\begin{tabular}{lllll}
                                      & $\beta$ & $\alpha$ & $\nu_{\perp}$ & $z$  \\
CDP in 2D~\cite{henkel2008non} & 0.64    & 0.42     & 0.80          & 1.53 \\
2D plane ($\phi_c=0.405(1)$)          & 0.63    & 0.42     & 0.80          & 1.53 \\
spherical surface ($\phi_c=0.405(1)$) & 0.66    & 0.42     & 0.89          & 1.53
\end{tabular}
\end{ruledtabular}
\end{table}

\begin{figure}
    \centering
    \includegraphics[width=0.95\textwidth]{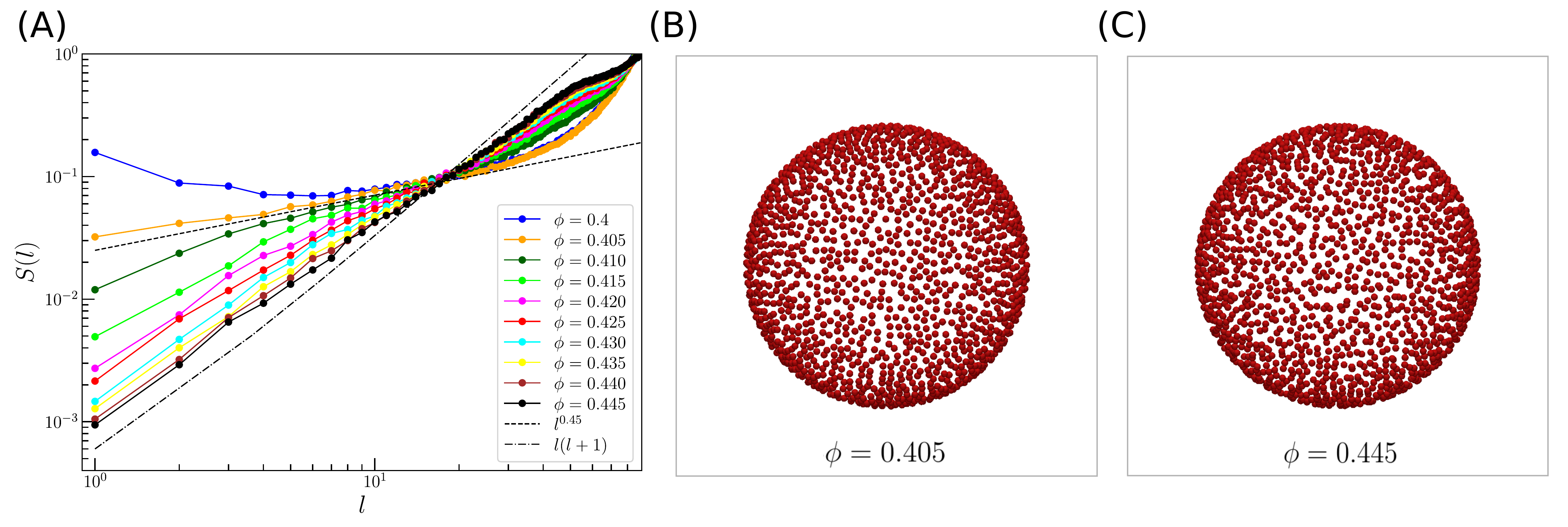}
    \caption{(A) Spherical structure factor $S(l)$ for the system of size ratio $\sigma/R=0.05$ at various packing fraction, {in which each $S(l)$ curve is averaged from 100 configurations obtained by independent simulations from random initial configurations.} {(B) A typical snapshot of hyperuniform particle distribution on sphere at critical point $\phi=0.405$. (C) A typical snapshot of hyperuniform particle distribution on sphere in active state $\phi=0.445$.}}
    \label{Fig.3}
\end{figure}

Next, we investigate the structural change in the absorbing transition. The structure factor of a system consisting of $N$ particles on a spherical surface of radius $R$ is defined as~\cite{bovzivc2019spherical,meyra2019hyperuniformity} 
\begin{equation}
S(l) = \frac{1}{N} \sum_{i,j = 1}^{N} P_l \left[ \cos \left( \frac{d_{ij}}{R}\right)\right],
\end{equation}
where $d_{ij}$ is the (great circle) spherical distance between particles $i$ and $j$, and $P_l(\cdot)$ is the $l$th order Legendre polynomial. Here $l$ is the wave number, which plays the role of wave vector as in Euclidean spaces. We plot $S(l)$ for systems of $\sigma/ R = 0.05$ at various packing fraction $\phi$ in Fig.~\ref{Fig.3}, and the absorbing transition occurs at $\phi_c \approx 0.405$. One can see that when $\phi > \phi_c$, $S(l) \rightarrow 0$ with $l \rightarrow 0$, which indicates that the system is hyperuniform~\cite{bovzivc2019spherical,meyra2019hyperuniformity}. At $\phi \approx \phi_c$, $S(l \rightarrow 0) \sim l^{0.45}$, which does not change with $R$~\cite{suppMat}. This is interesting, as in the random organization in 2D, at the absorbing transition point, $S(q \rightarrow 0) \sim q^{0.45}$ with $q$ the wave vector, which is called the critical hyperuniformity. Our results suggest that the critical hyperuniformity in random organization also exists in the system on a spherical surface, and the exponent in the hyperuniform scaling remains the same. 
Moreover, as shown in Fig.~\ref{Fig.3}, with increasing $\phi$, the exponent in the hyperuniform scaling increases and saturates in the active state with $S(l \rightarrow 0) \sim l(l+1)$, of which the number density fluctuation can be found in ~\cite{suppMat}. This is similar with the random organization in 2D, while the hyperuniform scaling in the non-equilibrium hyperuniform fluid in 2D is $S(q \rightarrow 0) \sim q^2$~\cite{lei2019nonequilibrium,lei2019hydrodynamics}.


To understand the hyperuniform scaling of the active state, we formulate a dynamic mean field theory for the active hyperuniform fluid on a spherical surface. Since the local dynamics of the system on the spherical surface is the same as the one in the flat 2D plane, we can extend the diffusion-like field equation in 2D to a spherical surface by preserving the general form and transferring directly into curvilinear coordinates $\mathbf{r}(\theta,\phi)$ for a 2D spherical surface~\cite{hexner2017noise,lei2019nonequilibrium}.

The dynamical field equation of the particle number density field $\rho(\mathbf{r}(\theta,\phi),t)$ on a sphere can be written as
\begin{equation}\label{eqdiff}
\frac{\partial \rho(\mathbf{r},t)}{\partial t}=D_0 \nabla^2 \rho(\mathbf{r},t)+\xi(\mathbf{r},t),
\end{equation}
where $D_0$ is the intrinsic diffusion coefficient and {the noise term $\xi(\mathbf{r},t)$ conserves the center of mass of the system, i.e., $\xi(\mathbf{r},t)=\sqrt{\bar{\rho}} \, \nabla^2 \eta(\mathbf{r},t)$~\cite{hexner2017noise}}.
We note that, the Laplacian Operator $\nabla^2$ here is in curvilinear form:
\begin{equation}
\nabla^{2} f=\frac{1}{R^{2} \sin \theta} \frac{\partial}{\partial \theta}\left(\sin \theta \frac{\partial f}{\partial \theta}\right)+\frac{1}{R^{2} \sin ^{2} \theta} \frac{\partial^{2} f}{\partial \varphi^{2}},
\end{equation}
where $\eta(\mathbf{r},t)$ is the uncorrelated Gaussian white noise, and $\left\langle\eta(\mathbf{r}, t) \eta\left(\mathbf{r}^{\prime}, t^{\prime}\right)\right\rangle=A^{2} \delta\left(\mathbf{r}-\mathbf{r}^{\prime}\right) \delta\left(t-t^{\prime}\right)$ with $A$ is the prefactor indicating the the noise strength.

We define the spherical Fourier transform in both temporal and spatial coordinate on local density $\rho(\mathbf{r},t)$ and noise $\eta(\mathbf{r},t)$ as
\begin{equation}
\rho_{l, \omega} = \iint \sin\theta d \theta d\phi \,  Y_l^0(\theta,\phi) \int d t e^{-i \omega t}\rho(\mathbf{r},t),
\end{equation}
\begin{equation}
\eta_{l, \omega} =\iint \sin\theta d \theta d\phi \,  Y_l^0(\theta,\phi) \int d t e^{-i \omega t} \eta(\mathbf{r},t).
\end{equation}
where the zonal spherical harmonics $Y_l^0$ satisfies $\nabla^2Y_l^0=-\l(l+1)/R^2 Y_l^0$. The homogeneous state, i.e., $\rho(\mathbf{r},t) = \bar{\rho}$, is the solution to Eq.~\ref{eqdiff}. By making small perturbation $\rho(\mathbf{r},t) = \bar{\rho} + \delta \rho(\mathbf{r},t)$ with the mathematical relation in~\cite{suppMat}, one can obtain a linearized equation in the Fourier space $(\omega, l)$ 
\begin{equation}
i \omega \delta \rho_{l, \omega}R^2=-D_0 l(l+1) \delta \rho_{l, \omega}-l(l+1) \sqrt{\bar{\rho}} \, \eta_{l, \omega}.
\end{equation}
Thus, the spectrum of density fluctuation of the particles is
\begin{equation}
\delta \rho_{l, \omega}=-\frac{l(l+1) \sqrt{\bar{\rho}} \, \eta_{l, \omega}}{i \omega R^2+D_0 l(l+1)},
\end{equation}
with which we obtain the dynamical spherical structure factor 
\begin{equation}
\begin{aligned}
S(l, \omega) &=\frac{1}{ 2 \pi \tau_{\max } N}\left\langle\delta \rho_{l, \omega} \delta \rho_{l, \omega}^{*}\right\rangle \\
&=\left(\frac{l^2(l+1)^2 \bar{\rho}}{\omega^{2}R^4+D_0^{2} l^2(l+1)^2}\right) \frac{1}{2 \pi \tau_{\max } N}\left\langle\eta_{l, \omega} \eta_{l, \omega}^{*}\right\rangle \\
&=\left(\frac{l^2(l+1)^2}{\omega^{2}R^4+D_0^{2} l^2(l+1)^2}\right) \frac{A^{2}}{2 \pi}.
\end{aligned}
\end{equation}
Here $\tau_{\max}$ is the maximum observation time defined by Fourier transform in the time scale $\int d t=\lim _{\tau_{\max } \rightarrow \infty} \int_{-\tau_{\max } / 2}^{+\tau_{\max } / 2} d t$ \cite{lei2019nonequilibrium}.
Then the static spherical structure factor is
\begin{equation}
S(l) =\int S(l, \omega) d \omega =\frac{A^{2}}{2 D_0 } \frac{l(l+1)}{R^2 } \sim  \frac{l(l+1)}{R^2},
\end{equation}
which agrees with the structure factor obtained in computer simulations in Fig.~\ref{Fig.3}. 

The mean field theory above describes systems with overdamped Langevin dynamics, in which the inertia of particles is neglected. However, in realistic fluids, e.g., active circle swimmers~\cite{zhang2022hyperuniform,huang2021circular}, each particle has a finite inertia, and the noise induced by the interaction between particles does not conserve the center of mass of the system but rather the momentum of the system. It was previously found that the in Euclidean spaces, the inertia of particles in non-equilibrium hyperuniform fluids sets a lengthscale $q_{HU}$, and the hyperuniform scaling $S(q) \sim q^2$ only exists at $q \ll q_{HU}$~\cite{lei2019hydrodynamics}. Here we investigate the effect of inertia on the non-equilibrium hyperuniform fluids on a sphere. To this end, we extend our dynamical field theory by replacing the center-of-mass conserving noise $\xi$ in Eq.~\ref{eqdiff} with an momentum-conserving noise that accounts for the inertia effect in underdamped systems. The mean field equations for a random organizing hard-sphere fluid on a spherical surface with finite inertia can be written as
\begin{equation}\label{eqdiffinertia}
\begin{aligned}
\frac{\partial\rho(\mathbf{r},t)}{\partial t}&=D_0\nabla^2\rho(\mathbf{r},t)+\nabla\cdot \mathbf{j}(\mathbf{r},t)\\
\frac{\partial\mathbf{j}(\mathbf{r},t)}{\partial t}&=-\gamma\mathbf{j}(\mathbf{r},t)+\gamma\sqrt{\bar{\rho}}\nabla\eta(\mathbf{r},t)
\end{aligned}
\end{equation}
where $\mathbf{j}$ is the momentum-conserving noise flux, and $\gamma$ is the friction coefficient with $\eta$ a Gaussian white noise. We can see that at the overdamped limit $\gamma \rightarrow \infty$, Eq.~\ref{eqdiffinertia} recovers Eq.~\ref{eqdiff}. By performing spherical Fourier transform and linear response around the homogeneous state, the dynamic field equation in the Fourier space can be written as
\begin{equation}\label{lineardiff}
\begin{aligned}
(i\omega)^2\delta\rho_{l,\omega}&=-\left(\gamma+D_0\frac{l(l+1)}{R^2}\right)i\omega\delta\rho_{l,\omega} \\
&-\gamma D_0 \frac{l(l+1)}{R^2} \delta\rho_{l,\omega} -\gamma\sqrt{\bar{\rho}}\frac{l(l+1)}{R^2}\eta_{l,\omega}.
\end{aligned}
\end{equation}
We assume that there is an effective length-scale dependent friction coefficient $\gamma_{\rm eff}=\gamma+D_0l(l+1)/R^2$, and we can rewrite Eq.~\ref{lineardiff} in $l$ space as 
\begin{equation}
\frac{1}{\gamma_{\mathrm{eff}}}\partial^2_t\delta\rho_{l}=-\,\partial_t\delta\rho_{l}-D_{\mathrm{eff}} \,\frac{l(l+1)}{R^2} \delta\rho_{l} -\alpha \sqrt{\bar{\rho}}\frac{l(l+1)}{R^2}\eta_{l},
\end{equation}
where the effective diffusion coefficient $D_{\mathrm{eff}}=\alpha D_0=\gamma D_0/(\gamma +D_0l(l+1)/R^2)$, and $\alpha=\gamma/\gamma_{\mathrm{eff}}$. At the overdamped limit $\gamma \rightarrow \infty$ or at the large lengthscale $l\rightarrow 0$, we have $\alpha \rightarrow 1$, $\gamma_{\mathrm{eff}} \rightarrow \gamma$, $D_{\mathrm{eff}} \rightarrow D_0$ and the noise strength $\alpha\sqrt{\bar{\rho}} \rightarrow \sqrt{\bar{\rho}}$. This suggests that the inertia affect disappears at the large length-scale $l \rightarrow 0$.

By solving Eq.~\ref{lineardiff} in Fourier space we obtain the dynamical spherical structure factor 
\begin{equation}
\begin{aligned}
S(l,\omega)&=\frac{1}{2\pi\tau_{max}}\left< \delta\rho_{l,\omega,} \delta \rho_{l,\omega}^* \right>\\
&=\frac{A\frac{l^2(l+1^2)}{R^4}\gamma^2}{2\pi(\omega^2+D_0^2\frac{l^2(l+1^2)}{R^4})(\omega^2+\gamma^2)}
\end{aligned},
\end{equation}
with which, we can obtain the static spherical structure factor 
\begin{equation}
\begin{aligned}
S(l)&= \int S(l,\omega)d\omega\\
&=\frac{A^{2} \gamma^{2} \frac{l^2(l+1)^2}{R^4}}{2 \pi\left(\gamma^{2}-D_0^{2} \frac{l^2(l+1)^2}{R^4}\right)}\left(\frac{\pi}{D_0 \frac{l^2(l+1)^2}{R^4}}-\frac{\pi}{\gamma}\right) \\
&=\frac{A^{2} \gamma \frac{l(l+1)}{R^2}}{2 D_0\left(\gamma+D_0 \frac{l(l+1)}{R^2}\right)}\\
&= \alpha \frac{ A^2 }{2D_0}\frac{l(l+1)}{R^2}.
\end{aligned}
\end{equation}
This suggests that similar to the non-equilibrium hyperuniform fluids in Euclidean spaces~\cite{lei2019hydrodynamics}, the inertia only sets a length-scale $l_{\rm inertia} \sim \sqrt{\gamma/D_0}$, above which the hyperuniform scaling $S(l) \sim l(l+1)/R^2$ appears.

In conclusion, we have investigated the absorbing transition and the random organizing hyperuniform fluids on a closed manifold, i.e., a spherical surface.
We find that the topological change of the space does not qualitatively affect the critical behaviour and hyperuniformity of random organizing systems. Based on a generalized definition of hyperuniformity on a spherical surface, i.e., spherical hyperuniformity, we numerically measure the spherical structure factor $S(l)$ and the corresponding number fluctuation $\sigma_n^2(a)$ within a window length-scale $a$ in the system. Our results show that the spherical hyperuniformity occurs at both the absorbing transition critical point and the active state on a spherical surface, with the spherical structure factor scalings $S(l \rightarrow 0)\rightarrow (l/R)^{0.45}$ and $S(l \rightarrow 0)\rightarrow l(l+1)/R^{2}$, respectively, which are similar to the random organizing systems in Euclidean spaces. To understand this, we extend the dynamical mean field theory for random organizing active state to curved space and derive the scaling of spherical structure factor $S(l \rightarrow 0 ) \sim l(l+1)/R^{2}$, which agrees quantitatively with computer simulations. We also find that the center-of-mass conserving noise is the main cause driving the formation of a dynamic hyperuniform state on a sphere. Furthermore, we discuss the effect of inertia in realistic particle systems on dynamic hyperuniformity by extending our field equation to underdamped systems. We find that the dynamic equation in Fourier space can be written with length-scale dependent diffusion and friction coefficients, which shows that the effect of inertia should disappear above certain length-scale, and at the overdamped limit or the large-scale limit, the scaling of structure factor recovers $S(l) \sim l(l+1)/R^2$, which is similar to the situation in Euclidean spaces~\cite{lei2019hydrodynamics}.

In addition to structural feature, we also investigate the dynamical criticality of random organization on a sphere. By treating the radius of the spherical surface as the size of the system, we perform a finite-size scaling analysis to determine several dynamical critical exponents of the absorbing-phase transition. Our results show that the obtained critical exponents are close to the value obtained in the 2D Euclidean space, which suggests that the absorbing transition on a sphere also belongs to the CDP universality class.
{Generally, our results suggest that the topology of the space may not qualitatively affect the phase behavior, criticality, and hyperuniformity in random organization.} It is noteworthy that the hyperuniform structures are long-range correlated~\cite{torquato2003local,torquato2018hyperuniform}, while the reciprocal excitation between particles are local dynamic rules, and it is non-trivial that the emergent long-range hyperuniform correlation does not depend on the topology of the space. Our work suggests a way to produce dynamically self-organized hyperuniform structures on a closed manifold without introducing interfaces, which normally has significant impacts in long-range correlated states. This may be useful for fabricating disordered photonic structures on curved spaces or hyperuniform receptor pattens for biosensing~\cite{xia2022receptor}.
Moreover, our finding indicates that random organizing dynamic hyperuniform states can maintain their hyperuniformity even when there are topological changes in space. This suggests that these dynamic hyperuniform states have the ability to self-repair against significant distortions in space. {A number of interesting questions are yet to be studied, e.g., effects of noises, particle shape and size polydispersity on the random organization on curved spaces~\cite{mapre2019,hexner2017enhanced}.}


\section*{Acknowledgements}
This work has been supported by the Singapore Ministry of Education through the Academic Research Fund MOE2019-T2-2-010. We thank NSCC for granting computational resources.

\section*{Supplementary Material}
See the supplementary material for the derivation of spherical Fourier transform and the simulation results for spherical surface of different sizes.

\section*{Data Availability}
The data that support the findings of this study are available within the article and its supplementary material and from the corresponding author upon reasonable request.

\bibliography{ref}

\begin{thebibliography}{35}
\expandafter\ifx\csname natexlab\endcsname\relax\def\natexlab#1{#1}\fi
\expandafter\ifx\csname bibnamefont\endcsname\relax
  \def\bibnamefont#1{#1}\fi
\expandafter\ifx\csname bibfnamefont\endcsname\relax
  \def\bibfnamefont#1{#1}\fi
\expandafter\ifx\csname citenamefont\endcsname\relax
  \def\citenamefont#1{#1}\fi
\expandafter\ifx\csname url\endcsname\relax
  \def\url#1{\texttt{#1}}\fi
\expandafter\ifx\csname urlprefix\endcsname\relax\def\urlprefix{URL }\fi
\providecommand{\bibinfo}[2]{#2}
\providecommand{\eprint}[2][]{\url{#2}}

\bibitem[{\citenamefont{Torquato and Stillinger}(2003)}]{torquato2003local}
\bibinfo{author}{\bibfnamefont{S.}~\bibnamefont{Torquato}} \bibnamefont{and}
  \bibinfo{author}{\bibfnamefont{F.~H.} \bibnamefont{Stillinger}},
  \bibinfo{journal}{Phys. Rev. E} \textbf{\bibinfo{volume}{68}},
  \bibinfo{pages}{041113} (\bibinfo{year}{2003}).

\bibitem[{\citenamefont{O{\u{g}}uz et~al.}(2017)\citenamefont{O{\u{g}}uz,
  Socolar, Steinhardt, and Torquato}}]{ouguz2017hyperuniformity}
\bibinfo{author}{\bibfnamefont{E.~C.} \bibnamefont{O{\u{g}}uz}},
  \bibinfo{author}{\bibfnamefont{J.~E.} \bibnamefont{Socolar}},
  \bibinfo{author}{\bibfnamefont{P.~J.} \bibnamefont{Steinhardt}},
  \bibnamefont{and} \bibinfo{author}{\bibfnamefont{S.}~\bibnamefont{Torquato}},
  \bibinfo{journal}{Phys. Rev. B} \textbf{\bibinfo{volume}{95}},
  \bibinfo{pages}{054119} (\bibinfo{year}{2017}).

\bibitem[{\citenamefont{Kim and Torquato}(2018)}]{kim2018effect}
\bibinfo{author}{\bibfnamefont{J.}~\bibnamefont{Kim}} \bibnamefont{and}
  \bibinfo{author}{\bibfnamefont{S.}~\bibnamefont{Torquato}},
  \bibinfo{journal}{Phys. Rev. B} \textbf{\bibinfo{volume}{97}},
  \bibinfo{pages}{054105} (\bibinfo{year}{2018}).

\bibitem[{\citenamefont{Lomba et~al.}(2017)\citenamefont{Lomba, Weis, and
  Torquato}}]{lomba2017disordered}
\bibinfo{author}{\bibfnamefont{E.}~\bibnamefont{Lomba}},
  \bibinfo{author}{\bibfnamefont{J.-J.} \bibnamefont{Weis}}, \bibnamefont{and}
  \bibinfo{author}{\bibfnamefont{S.}~\bibnamefont{Torquato}},
  \bibinfo{journal}{Phys. Rev. E} \textbf{\bibinfo{volume}{96}},
  \bibinfo{pages}{062126} (\bibinfo{year}{2017}).

\bibitem[{\citenamefont{Zhang et~al.}(2016)\citenamefont{Zhang, Stillinger, and
  Torquato}}]{zhang2016perfect}
\bibinfo{author}{\bibfnamefont{G.}~\bibnamefont{Zhang}},
  \bibinfo{author}{\bibfnamefont{F.~H.} \bibnamefont{Stillinger}},
  \bibnamefont{and} \bibinfo{author}{\bibfnamefont{S.}~\bibnamefont{Torquato}},
  \bibinfo{journal}{Sci. Rep.} \textbf{\bibinfo{volume}{6}}, \bibinfo{pages}{1}
  (\bibinfo{year}{2016}).

\bibitem[{\citenamefont{Donev et~al.}(2005)\citenamefont{Donev, Stillinger, and
  Torquato}}]{donev2005unexpected}
\bibinfo{author}{\bibfnamefont{A.}~\bibnamefont{Donev}},
  \bibinfo{author}{\bibfnamefont{F.~H.} \bibnamefont{Stillinger}},
  \bibnamefont{and} \bibinfo{author}{\bibfnamefont{S.}~\bibnamefont{Torquato}},
  \bibinfo{journal}{Phys. Rev. Lett.} \textbf{\bibinfo{volume}{95}},
  \bibinfo{pages}{090604} (\bibinfo{year}{2005}).

\bibitem[{\citenamefont{Ricouvier et~al.}(2017)\citenamefont{Ricouvier,
  Pierrat, Carminati, Tabeling, and Yazhgur}}]{ricouvier2017optimizing}
\bibinfo{author}{\bibfnamefont{J.}~\bibnamefont{Ricouvier}},
  \bibinfo{author}{\bibfnamefont{R.}~\bibnamefont{Pierrat}},
  \bibinfo{author}{\bibfnamefont{R.}~\bibnamefont{Carminati}},
  \bibinfo{author}{\bibfnamefont{P.}~\bibnamefont{Tabeling}}, \bibnamefont{and}
  \bibinfo{author}{\bibfnamefont{P.}~\bibnamefont{Yazhgur}},
  \bibinfo{journal}{Phys. Rev. Lett.} \textbf{\bibinfo{volume}{119}},
  \bibinfo{pages}{208001} (\bibinfo{year}{2017}).

\bibitem[{\citenamefont{Jiao et~al.}(2014)\citenamefont{Jiao, Lau, Hatzikirou,
  Meyer-Hermann, Corbo, and Torquato}}]{jiao2014avian}
\bibinfo{author}{\bibfnamefont{Y.}~\bibnamefont{Jiao}},
  \bibinfo{author}{\bibfnamefont{T.}~\bibnamefont{Lau}},
  \bibinfo{author}{\bibfnamefont{H.}~\bibnamefont{Hatzikirou}},
  \bibinfo{author}{\bibfnamefont{M.}~\bibnamefont{Meyer-Hermann}},
  \bibinfo{author}{\bibfnamefont{J.~C.} \bibnamefont{Corbo}}, \bibnamefont{and}
  \bibinfo{author}{\bibfnamefont{S.}~\bibnamefont{Torquato}},
  \bibinfo{journal}{Phys. Rev. E} \textbf{\bibinfo{volume}{89}},
  \bibinfo{pages}{022721} (\bibinfo{year}{2014}).

\bibitem[{\citenamefont{Zheng et~al.}(2020)\citenamefont{Zheng, Li, and
  Ciamarra}}]{zheng2020hyperuniformity}
\bibinfo{author}{\bibfnamefont{Y.}~\bibnamefont{Zheng}},
  \bibinfo{author}{\bibfnamefont{Y.-W.} \bibnamefont{Li}}, \bibnamefont{and}
  \bibinfo{author}{\bibfnamefont{M.~P.} \bibnamefont{Ciamarra}},
  \bibinfo{journal}{Soft Matter} \textbf{\bibinfo{volume}{16}},
  \bibinfo{pages}{5942} (\bibinfo{year}{2020}).

\bibitem[{\citenamefont{Weijs et~al.}(2015)\citenamefont{Weijs, Jeanneret,
  Dreyfus, and Bartolo}}]{weijs2015emergent}
\bibinfo{author}{\bibfnamefont{J.~H.} \bibnamefont{Weijs}},
  \bibinfo{author}{\bibfnamefont{R.}~\bibnamefont{Jeanneret}},
  \bibinfo{author}{\bibfnamefont{R.}~\bibnamefont{Dreyfus}}, \bibnamefont{and}
  \bibinfo{author}{\bibfnamefont{D.}~\bibnamefont{Bartolo}},
  \bibinfo{journal}{Phys. Rev. Lett.} \textbf{\bibinfo{volume}{115}},
  \bibinfo{pages}{108301} (\bibinfo{year}{2015}).

\bibitem[{\citenamefont{Castillo et~al.}(2019)\citenamefont{Castillo, Mujica,
  Sep{\'u}lveda, Sobarzo, Guzm{\'a}n, and Soto}}]{castillo2019hyperuniform}
\bibinfo{author}{\bibfnamefont{G.}~\bibnamefont{Castillo}},
  \bibinfo{author}{\bibfnamefont{N.}~\bibnamefont{Mujica}},
  \bibinfo{author}{\bibfnamefont{N.}~\bibnamefont{Sep{\'u}lveda}},
  \bibinfo{author}{\bibfnamefont{J.~C.} \bibnamefont{Sobarzo}},
  \bibinfo{author}{\bibfnamefont{M.}~\bibnamefont{Guzm{\'a}n}},
  \bibnamefont{and} \bibinfo{author}{\bibfnamefont{R.}~\bibnamefont{Soto}},
  \bibinfo{journal}{Phys. Rev. E} \textbf{\bibinfo{volume}{100}},
  \bibinfo{pages}{032902} (\bibinfo{year}{2019}).

\bibitem[{\citenamefont{Mitra et~al.}(2021)\citenamefont{Mitra, Parmar,
  Leishangthem, Sastry, and Foffi}}]{Mitra_2021}
\bibinfo{author}{\bibfnamefont{S.}~\bibnamefont{Mitra}},
  \bibinfo{author}{\bibfnamefont{A.~D.~S.} \bibnamefont{Parmar}},
  \bibinfo{author}{\bibfnamefont{P.}~\bibnamefont{Leishangthem}},
  \bibinfo{author}{\bibfnamefont{S.}~\bibnamefont{Sastry}}, \bibnamefont{and}
  \bibinfo{author}{\bibfnamefont{G.}~\bibnamefont{Foffi}}, \bibinfo{journal}{J.
  Stat. Mech.: Theory Exp} \textbf{\bibinfo{volume}{2021}},
  \bibinfo{pages}{033203} (\bibinfo{year}{2021}).

\bibitem[{\citenamefont{Hexner and Levine}(2015)}]{hexner2015hyperuniformity}
\bibinfo{author}{\bibfnamefont{D.}~\bibnamefont{Hexner}} \bibnamefont{and}
  \bibinfo{author}{\bibfnamefont{D.}~\bibnamefont{Levine}},
  \bibinfo{journal}{Phys. Rev. Lett.} \textbf{\bibinfo{volume}{114}},
  \bibinfo{pages}{110602} (\bibinfo{year}{2015}).

\bibitem[{\citenamefont{Tjhung and Berthier}(2015)}]{tjhung2015hyperuniform}
\bibinfo{author}{\bibfnamefont{E.}~\bibnamefont{Tjhung}} \bibnamefont{and}
  \bibinfo{author}{\bibfnamefont{L.}~\bibnamefont{Berthier}},
  \bibinfo{journal}{Phys. Rev. Lett.} \textbf{\bibinfo{volume}{114}},
  \bibinfo{pages}{148301} (\bibinfo{year}{2015}).

\bibitem[{\citenamefont{Hexner and Levine}(2017)}]{hexner2017noise}
\bibinfo{author}{\bibfnamefont{D.}~\bibnamefont{Hexner}} \bibnamefont{and}
  \bibinfo{author}{\bibfnamefont{D.}~\bibnamefont{Levine}},
  \bibinfo{journal}{Phys. Rev. Lett.} \textbf{\bibinfo{volume}{118}},
  \bibinfo{pages}{020601} (\bibinfo{year}{2017}).

\bibitem[{\citenamefont{Hexner et~al.}(2017)\citenamefont{Hexner, Chaikin, and
  Levine}}]{hexner2017enhanced}
\bibinfo{author}{\bibfnamefont{D.}~\bibnamefont{Hexner}},
  \bibinfo{author}{\bibfnamefont{P.~M.} \bibnamefont{Chaikin}},
  \bibnamefont{and} \bibinfo{author}{\bibfnamefont{D.}~\bibnamefont{Levine}},
  \bibinfo{journal}{Proc. Natl Acad. Sci. USA} \textbf{\bibinfo{volume}{114}},
  \bibinfo{pages}{4294} (\bibinfo{year}{2017}).

\bibitem[{\citenamefont{Wang et~al.}(2018)\citenamefont{Wang, Schwarz, and
  Paulsen}}]{wang2018hyperuniformity}
\bibinfo{author}{\bibfnamefont{J.}~\bibnamefont{Wang}},
  \bibinfo{author}{\bibfnamefont{J.~M.} \bibnamefont{Schwarz}},
  \bibnamefont{and} \bibinfo{author}{\bibfnamefont{J.~D.}
  \bibnamefont{Paulsen}}, \bibinfo{journal}{Nat. Commun.}
  \textbf{\bibinfo{volume}{9}}, \bibinfo{pages}{1} (\bibinfo{year}{2018}).

\bibitem[{\citenamefont{Wilken et~al.}(2020)\citenamefont{Wilken, Guerra, Pine,
  and Chaikin}}]{wilken2020hyperuniform}
\bibinfo{author}{\bibfnamefont{S.}~\bibnamefont{Wilken}},
  \bibinfo{author}{\bibfnamefont{R.~E.} \bibnamefont{Guerra}},
  \bibinfo{author}{\bibfnamefont{D.~J.} \bibnamefont{Pine}}, \bibnamefont{and}
  \bibinfo{author}{\bibfnamefont{P.~M.} \bibnamefont{Chaikin}},
  \bibinfo{journal}{Phys. Rev. Lett.} \textbf{\bibinfo{volume}{125}},
  \bibinfo{pages}{148001} (\bibinfo{year}{2020}).

\bibitem[{\citenamefont{Lei et~al.}(2019)\citenamefont{Lei, Ciamarra, and
  Ni}}]{lei2019nonequilibrium}
\bibinfo{author}{\bibfnamefont{Q.-L.} \bibnamefont{Lei}},
  \bibinfo{author}{\bibfnamefont{M.~P.} \bibnamefont{Ciamarra}},
  \bibnamefont{and} \bibinfo{author}{\bibfnamefont{R.}~\bibnamefont{Ni}},
  \bibinfo{journal}{Sci. Adv.} \textbf{\bibinfo{volume}{5}},
  \bibinfo{pages}{eaau7423} (\bibinfo{year}{2019}).

\bibitem[{\citenamefont{Lei and Ni}(2019)}]{lei2019hydrodynamics}
\bibinfo{author}{\bibfnamefont{Q.-L.} \bibnamefont{Lei}} \bibnamefont{and}
  \bibinfo{author}{\bibfnamefont{R.}~\bibnamefont{Ni}}, \bibinfo{journal}{Proc.
  Natl. Acad. Sci. USA} \textbf{\bibinfo{volume}{116}}, \bibinfo{pages}{22983}
  (\bibinfo{year}{2019}).

\bibitem[{\citenamefont{Zhang and Snezhko}(2022)}]{zhang2022hyperuniform}
\bibinfo{author}{\bibfnamefont{B.}~\bibnamefont{Zhang}} \bibnamefont{and}
  \bibinfo{author}{\bibfnamefont{A.}~\bibnamefont{Snezhko}},
  \bibinfo{journal}{Phys. Rev. Lett.} \textbf{\bibinfo{volume}{128}},
  \bibinfo{pages}{218002} (\bibinfo{year}{2022}).

\bibitem[{\citenamefont{Huang et~al.}(2021)\citenamefont{Huang, Hu, Yang, Liu,
  and Zhang}}]{huang2021circular}
\bibinfo{author}{\bibfnamefont{M.}~\bibnamefont{Huang}},
  \bibinfo{author}{\bibfnamefont{W.}~\bibnamefont{Hu}},
  \bibinfo{author}{\bibfnamefont{S.}~\bibnamefont{Yang}},
  \bibinfo{author}{\bibfnamefont{Q.-X.} \bibnamefont{Liu}}, \bibnamefont{and}
  \bibinfo{author}{\bibfnamefont{H.}~\bibnamefont{Zhang}},
  \bibinfo{journal}{Proc. Natl. Acad. Sci. USA} \textbf{\bibinfo{volume}{118}},
  \bibinfo{pages}{e2100493118} (\bibinfo{year}{2021}).

\bibitem[{\citenamefont{Xia et~al.}(2023)\citenamefont{Xia, Zhang, Ciamarra,
  Jiao, and Ni}}]{xia2022receptor}
\bibinfo{author}{\bibfnamefont{X.}~\bibnamefont{Xia}},
  \bibinfo{author}{\bibfnamefont{G.}~\bibnamefont{Zhang}},
  \bibinfo{author}{\bibfnamefont{M.~P.} \bibnamefont{Ciamarra}},
  \bibinfo{author}{\bibfnamefont{Y.}~\bibnamefont{Jiao}}, \bibnamefont{and}
  \bibinfo{author}{\bibfnamefont{R.}~\bibnamefont{Ni}}, \bibinfo{journal}{JACS
  Au} \textbf{\bibinfo{volume}{in press}} (\bibinfo{year}{2023}).

\bibitem[{\citenamefont{Florescu et~al.}(2009)\citenamefont{Florescu, Torquato,
  and Steinhardt}}]{florescu2009designer}
\bibinfo{author}{\bibfnamefont{M.}~\bibnamefont{Florescu}},
  \bibinfo{author}{\bibfnamefont{S.}~\bibnamefont{Torquato}}, \bibnamefont{and}
  \bibinfo{author}{\bibfnamefont{P.~J.} \bibnamefont{Steinhardt}},
  \bibinfo{journal}{Proc. Natl. Acad. Sci. USA} \textbf{\bibinfo{volume}{106}},
  \bibinfo{pages}{20658} (\bibinfo{year}{2009}).

\bibitem[{\citenamefont{Man et~al.}(2013)\citenamefont{Man, Florescu,
  Williamson, He, Hashemizad, Leung, Liner, Torquato, Chaikin, and
  Steinhardt}}]{man2013isotropic}
\bibinfo{author}{\bibfnamefont{W.}~\bibnamefont{Man}},
  \bibinfo{author}{\bibfnamefont{M.}~\bibnamefont{Florescu}},
  \bibinfo{author}{\bibfnamefont{E.~P.} \bibnamefont{Williamson}},
  \bibinfo{author}{\bibfnamefont{Y.}~\bibnamefont{He}},
  \bibinfo{author}{\bibfnamefont{S.~R.} \bibnamefont{Hashemizad}},
  \bibinfo{author}{\bibfnamefont{B.~Y.} \bibnamefont{Leung}},
  \bibinfo{author}{\bibfnamefont{D.~R.} \bibnamefont{Liner}},
  \bibinfo{author}{\bibfnamefont{S.}~\bibnamefont{Torquato}},
  \bibinfo{author}{\bibfnamefont{P.~M.} \bibnamefont{Chaikin}},
  \bibnamefont{and} \bibinfo{author}{\bibfnamefont{P.~J.}
  \bibnamefont{Steinhardt}}, \bibinfo{journal}{Proc. Natl Acad. Sci. USA}
  \textbf{\bibinfo{volume}{110}}, \bibinfo{pages}{15886}
  (\bibinfo{year}{2013}).

\bibitem[{\citenamefont{Leseur et~al.}(2016)\citenamefont{Leseur, Pierrat, and
  Carminati}}]{leseur2016high}
\bibinfo{author}{\bibfnamefont{O.}~\bibnamefont{Leseur}},
  \bibinfo{author}{\bibfnamefont{R.}~\bibnamefont{Pierrat}}, \bibnamefont{and}
  \bibinfo{author}{\bibfnamefont{R.}~\bibnamefont{Carminati}},
  \bibinfo{journal}{Optica} \textbf{\bibinfo{volume}{3}}, \bibinfo{pages}{763}
  (\bibinfo{year}{2016}).

\bibitem[{\citenamefont{Brauchart et~al.}(2019)\citenamefont{Brauchart,
  Grabner, and Kusner}}]{brauchart2019hyperuniform}
\bibinfo{author}{\bibfnamefont{J.~S.} \bibnamefont{Brauchart}},
  \bibinfo{author}{\bibfnamefont{P.~J.} \bibnamefont{Grabner}},
  \bibnamefont{and} \bibinfo{author}{\bibfnamefont{W.}~\bibnamefont{Kusner}},
  \bibinfo{journal}{Constructive approximation} \textbf{\bibinfo{volume}{50}},
  \bibinfo{pages}{45} (\bibinfo{year}{2019}).

\bibitem[{\citenamefont{Meyra et~al.}(2019)\citenamefont{Meyra,
  Zarragoicoechea, Maltz, Lomba, and Torquato}}]{meyra2019hyperuniformity}
\bibinfo{author}{\bibfnamefont{A.~G.} \bibnamefont{Meyra}},
  \bibinfo{author}{\bibfnamefont{G.~J.} \bibnamefont{Zarragoicoechea}},
  \bibinfo{author}{\bibfnamefont{A.~L.} \bibnamefont{Maltz}},
  \bibinfo{author}{\bibfnamefont{E.}~\bibnamefont{Lomba}}, \bibnamefont{and}
  \bibinfo{author}{\bibfnamefont{S.}~\bibnamefont{Torquato}},
  \bibinfo{journal}{Phys. Rev. E} \textbf{\bibinfo{volume}{100}},
  \bibinfo{pages}{022107} (\bibinfo{year}{2019}).

\bibitem[{\citenamefont{Bo{\v{z}}i{\v{c}} and
  {\v{C}}opar}(2019)}]{bovzivc2019spherical}
\bibinfo{author}{\bibfnamefont{A.~L.} \bibnamefont{Bo{\v{z}}i{\v{c}}}}
  \bibnamefont{and}
  \bibinfo{author}{\bibfnamefont{S.}~\bibnamefont{{\v{C}}opar}},
  \bibinfo{journal}{Phys. Rev. E} \textbf{\bibinfo{volume}{99}},
  \bibinfo{pages}{032601} (\bibinfo{year}{2019}).

\bibitem[{not()}]{note1}
\bibinfo{note}{Note that the random organization can be generalized to other
  curved spaces by adopting similar modifications. Basically, when we give
  active particles paired displacements, the random displacements are
  introduced in the tangent plane and projected back to the curved surface.
  Therefore, this ensures that the local dynamics remains the same when we
  consider the field description of the system.}

\bibitem[{\citenamefont{Rossi et~al.}(2000)\citenamefont{Rossi,
  Pastor-Satorras, and Vespignani}}]{rossi2000universality}
\bibinfo{author}{\bibfnamefont{M.}~\bibnamefont{Rossi}},
  \bibinfo{author}{\bibfnamefont{R.}~\bibnamefont{Pastor-Satorras}},
  \bibnamefont{and}
  \bibinfo{author}{\bibfnamefont{A.}~\bibnamefont{Vespignani}},
  \bibinfo{journal}{Phys. Rev. Lett.} \textbf{\bibinfo{volume}{85}},
  \bibinfo{pages}{1803} (\bibinfo{year}{2000}).

\bibitem[{\citenamefont{Henkel et~al.}(2008)\citenamefont{Henkel, Hinrichsen,
  L{\"u}beck, and Pleimling}}]{henkel2008non}
\bibinfo{author}{\bibfnamefont{M.}~\bibnamefont{Henkel}},
  \bibinfo{author}{\bibfnamefont{H.}~\bibnamefont{Hinrichsen}},
  \bibinfo{author}{\bibfnamefont{S.}~\bibnamefont{L{\"u}beck}},
  \bibnamefont{and}
  \bibinfo{author}{\bibfnamefont{M.}~\bibnamefont{Pleimling}},
  \emph{\bibinfo{title}{Non-equilibrium phase transitions}},
  vol.~\bibinfo{volume}{1} (\bibinfo{publisher}{Springer},
  \bibinfo{year}{2008}).

\bibitem[{sup()}]{suppMat}
\bibinfo{note}{See Supplemental Material}.

\bibitem[{\citenamefont{Torquato}(2018)}]{torquato2018hyperuniform}
\bibinfo{author}{\bibfnamefont{S.}~\bibnamefont{Torquato}},
  \bibinfo{journal}{Phys. Rep.} \textbf{\bibinfo{volume}{745}},
  \bibinfo{pages}{1} (\bibinfo{year}{2018}).

\bibitem[{\citenamefont{Ma and Torquato}(2019)}]{mapre2019}
\bibinfo{author}{\bibfnamefont{Z.}~\bibnamefont{Ma}} \bibnamefont{and}
  \bibinfo{author}{\bibfnamefont{S.}~\bibnamefont{Torquato}},
  \bibinfo{journal}{Phys. Rev. E} \textbf{\bibinfo{volume}{99}},
  \bibinfo{pages}{022115} (\bibinfo{year}{2019}).

\end{thebibliography}

\end{document}